\documentclass[draft,aps,showpacs]{revtex4}
\newcommand{\ii}{\'{\i}}
\newcommand{\bb}{$\beta\beta$}
\newcommand{\bt}{$\beta\beta_{2\nu}$}
\newcommand{\bz}{$\beta\beta_{0\nu}$}
\newcommand{\gd}{$^{160}$Gd}
\newcommand{\dy}{$^{160}$Dy}

\begin{document}

\title{Theoretical description of the double beta decay of \gd}
\author{Jorge G. Hirsch}
\email{hirsch@nuclecu.unam.mx}
\author{Octavio Casta\~nos}
\email{ocasta@nuclecu.unam.mx}
\author{Peter O. Hess}
\email{hess@nuclecu.unam.mx}
\affiliation{Instituto de Ciencias Nucleares, Universidad Nacional Aut\'onoma de
M\'exico, A. P. 70-543 M\'exico 04510 D.F.}
\author{Osvaldo Civitarese}
\email{civitare@venus.fisica.unlp.edu.ar}
\affiliation{Departamento de F\ii sica, Universidad Nacional de La Plata,
 c.c.67; 1900, La Plata, Argentina}

\begin{abstract}
The half-life of the \bt ~decay of $^{160}$Gd, a process which was
previously reported as theoretically forbidden in the context of
the pseudo SU(3) model, is estimated by including the pairing
interaction. Thus, different occupations can be mixed by the
interaction opening new channels for the decay. Explicit
expressions are presented for the mixing induced by the pairing
force. Matrix elements for the \bt ~and \bz ~decays are
calculated, in the present pseudo SU(3) approach, by assuming both
a dominant component in the wave function of the ground state of
$^{160}$Gd and a more general model space. The results, of the
calculated \bb ~half-lives, suggest that the planned experiments
would succeed in detecting the \bt ~decay of \gd ~ and,
eventually, would improve the limits for the zero neutrino mode.
\end{abstract}


\maketitle

\section{Introduction}

The flux of solar and atmospheric neutrinos has been measured with
increasing precision, and the data offer direct evidence on the
presence of neutrino oscillations \cite{SK01,SNO01}. These
findings imply that at least neutrinos of some flavor should be
massive.  The difference between the square of the masses of
neutrinos belonging to different families can be extracted from
the experimental data, relying on theoretical models of mass
hierarchies and textures \cite{Bah01}. On the other hand, their
absolute scale cannot be obtained from these experiments.

The neutrinoless double beta decay (\bz), if detected, would
provide the complementary information needed to determine neutrino
masses, and would also offer definitive evidence that the neutrino
is a Majorana particle, i.e. that it is its own antiparticle
\cite{Ver86,Suh98}.

Theoretical nuclear matrix elements are needed to translate
experimental half-life limits, which are available for many
$\beta\beta$-unstable isotopes \cite{Moe93,Bar01}, into constrains
for the effective Majorana mass of the neutrino and, eventually,
for the contribution of right-handed currents to the weak
interactions. Thus, these matrix elements are essential to
understand the underlying physics.

The two neutrino mode of the double beta decay (\bt ) is allowed
as a second order process in the standard model. It has been
detected in ten nuclei \cite{Moe93,Bar01} and it has served as a
test of a variety of nuclear models \cite{Suh98}. The calculation
of the \bt ~and $\beta\beta_{0\nu}$ matrix elements requires the
use of different theoretical methods. Therefore a successful
prediction of the former cannot be considered a rigorous test of
the latter \cite{Ell02}. However, in most cases it is the best
available proof we can impose to a nuclear model used to predict
the $\beta\beta_{0\nu}$ matrix elements. When possible, the test
should also include the calculation of the energy spectra,
electromagnetic transitions and particle transfer observables, in
the  neighborhood of the double beta emitters.

Many experimental groups have reported measurements of \bb ~
processes \cite{Moe93,Mor99}. In direct-counting experiments the
analysis of the sum-energy spectrum of the emitted electrons
identifies the different $\beta\beta$-decay modes \cite{Pie94}.

The pseudo SU(3) approach has been used to describe many low-lying
rotational bands, as well as BE(2) and B(M1) intensities in rare
earth and actinide nuclei, both with even-even and odd-mass
numbers \cite{Beu98,Rom98,Var00a,Beu00,Var00b,Pop00}. The
$\beta\beta$ half-lives of some of these parent  nuclei to the
ground and excited states of the daughter were evaluated for the
two and zero neutrino emitting modes
\cite{Cas94,Hir95a,Hir95b,Hir95c,Hir95d} using the pseudo SU(3)
scheme. The predictions were found to be in good agreement with
the available experimental data for $^{150}$Nd and $^{238}$U.

Based on the selection rules of the simplest pseudo SU(3) model,
the theory predicts the complete suppression of the \bb ~decay for
the following five nuclei: $^{154}$Sm, $^{160}$Gd, $^{176}$Yb,
$^{232}$Th and $^{244}$Pu \cite{Cas94}. It was expected that these
forbidden decays would have, in the best case, matrix elements
that would be no greater than $20\%$ of the allowed ones,
resulting in the increase, by at least one order of magnitude, of
the predicted half-life \cite{Hir95d}. Experimental limits for the
$\beta\beta$ decay of $^{160}$Gd have been reported
\cite{Bur95,Kob95}. Recently it was argued that the strong
cancellation of the 2$\nu$ mode in the $\beta\beta$ decay of
$^{160}$Gd would suppress the background for the detection of the
0$\nu$ mode \cite{Dan00}.

In the present contribution we extend the previous research
\cite{Cas94,Hir95a,Hir95b,Hir95c,Hir95d} and  evaluate the
$\beta\beta$ half-lives of $^{160}$Gd using the pseudo SU(3)
model. While the 2$\nu$ mode is forbidden when the most probable
occupations are considered, states with different occupation
numbers can be activated by the pairing interaction. The amount of
this mixing is evaluated, and the possibility of observing the
$\beta\beta$ decay in $^{160}$Gd is discussed for both the 2$\nu$
and 0$\nu$ modes. The analysis is performed firstly for the
extreme case of a single active configuration in the initial
nucleus, and afterwards, the results are compared with the ones
obtained by enlarging the model space.

The paper is organized as follows. In Section II the pseudo SU(3)
formalism and the model Hamiltonian are briefly reviewed. In
Section III the \gd ~and \dy ~ground state wave functions are
built. Section IV and V contain the explicit formulae needed to
evaluate, using the pseudo SU(3) scheme, the \bt ~ and \bz ~matrix
elements, respectively. In Section VI the \gd ~ \bb ~nuclear
matrix elements and half-lives are presented. Conclusions are
drawn in the last Section.

\section{ The pseudo SU(3) formalism}

In order to obtain a microscopic description of the ground states
of \gd ~and \dy ~we will use the pseudo SU(3) model, which
successfully  describes collective excitations in rare earth
nuclei and actinides
\cite{Dra84,Beu98,Rom98,Var00a,Beu00,Var00b,Pop00}, and it has
been used to evaluate the half-lives of the \bt ~decay to the
ground and excited states ~and $\beta\beta_{0\nu}$ decays of six
heavy deformed nuclei \cite{Cas94,Hir95a,Hir95b,Hir95c,Hir95d},
and the double electron capture decays in other three nuclei
\cite{Cer99}.

In the pseudo $SU(3)$ shell model coupling scheme \cite{Rat73}
normal parity orbitals with quantum numbers $(\eta ,l,j)$ are
mapped to orbitals of another harmonic oscillator with $(\tilde
\eta = \eta-1, \tilde l, \tilde j)$. This set of orbitals, with
$\tilde j = j = \tilde l + \tilde s$, pseudo spin $\tilde s =1/2$
and pseudo orbital angular momentum $\tilde l$, define the
so-called pseudo space. For configurations of identical particles
occupying a single j orbital of abnormal parity, a simple
characterization of states is made by means of the seniority
coupling scheme.

The first step in the pseudo SU(3) description of a given nucleus
consists in finding the occupation numbers for protons and
neutrons in the normal and abnormal parity states $n^N_\pi,
n^N_\nu, n^A_\pi, n^A_\nu$ \cite{Cas94}. These numbers are
determined filling the Nilsson levels from below, as discussed in
\cite{Cas94}.

For even-even heavy nuclei it has been shown that if the residual
neutron-proton interaction is of the quadrupole type, regardless
of the interaction in the proton and neutron spaces, the most
important normal parity configurations are those with highest
spatial symmetry $\{ \tilde f_\alpha \} = \{ 2^{n^N_\alpha /2}\}$
\cite{Dra84}. This statement is valid for yrast states below the
backbending region. It implies that $ \tilde S_\pi = \tilde S_\nu
= 0$, i.e. only pseudo spin zero configurations are taken into
account.

Additionally in the abnormal parity space only seniority zero
configurations are taken into account.  This simplification implies that
 $J^A_\pi = J^A_\nu = 0$. This is a very
strong assumption quite useful in order to simplify the calculations.

Many-particle states of $n_\alpha$ active nucleons in a given normal
parity shell $\eta_\alpha$, $\alpha = \nu$ or $\pi$, can be classified by
the following chains of groups:

\begin{eqnarray}
~ \{ 1^{n^{N}_\alpha} \} ~~~~~~ \{ \tilde{f}_\alpha \} ~~~~~\{ f_\alpha
\} ~\gamma_\alpha ~ (\lambda_\alpha , \mu_\alpha ) ~~~ \tilde{S}_\alpha
~~ K_\alpha  ~\tilde{L}_\alpha  ~~~~~~~~~~~~~~~~~~~~~ J^N_\alpha ~~~~ \nonumber \\
U(\Omega^N_\alpha ) \supset U(\Omega^N_\alpha / 2 ) \times U(2) \supset
SU(3) \times SU(2) \supset %\nonumber \\
SO(3) \times SU(2) \supset SU_J(2),
\label{eq:chains}
\end{eqnarray}

\noindent where above each group the quantum numbers that
characterize its irreps are given and $\gamma_\alpha$ and
$K_\alpha$ are multiplicity labels of the indicated reductions.
The most important configurations are those with the highest
spatial symmetry \cite{Dra84,Var98}, namely those with pseudo spin
zero.

The model Hamiltonian contains spherical Nilsson single-particle terms for protons
($H_{sp,\pi}$) and neutrons ($H_{sp,\nu}$), the quadrupole-quadrupole
($\tilde Q \cdot \tilde Q$) and pairing interactions ($V_{pair}$), as well as
three `rotor-like' terms which are diagonal in the SU(3) basis.

\begin{eqnarray}
 H  = H_{sp,\pi} + H_{sp,\nu} +~V_{pair}
         - \frac{1}{2}~ \chi~ \tilde Q \cdot \tilde Q
      + ~a~ K_J^2~ +~ b~ J^2~ +~ A_{asym}~ \hat C_2 .\label{hamil}
\end{eqnarray}

This Hamiltonian can be separated into two parts: the first one includes
Nilsson single-particle energies and the pairing and quadrupole-quadrupole
interactions ($\tilde{Q}$ is the quadrupole operator in the pseudo SU(3)
space).
They are the basic components of any realistic Hamiltonian \cite{Rin79,Duf96}
and have been widely studied in the nuclear physics literature, allowing
their respective strengths to be fixed by systematics \cite{Rin79,Duf96}.
In the second one there are three rotor terms used to fine tune the moment of
inertia and the position of the different $K$ bands. The SU(3) mixing is due to
the single-particle and pairing terms.

The three `rotor-like' terms have been studied in detail in
previous papers where the pseudo SU(3) symmetry was used as a
dynamical symmetry \cite{Dra84}. In recent works, $a$, $b$, and
$A_{asym}$ were the only parameters used to fit the spectra
\cite{Var00a,Var00b,Var01,Pop00,Dra01}.

The spherical single-particle Nilsson Hamiltonian is

\begin{equation}
H_{sp} = \hbar \omega_0 (\eta + \frac{3}{2}) - \kappa \hbar {\omega}_0 \{ 2
\vec{l} \cdot \vec{s} + \mu \vec{l}^2 \}
= \sum_i \epsilon (\eta_i ,l_i ,j_i ) a^\dagger_i a_i
\label{Nilssonh}
\end{equation}

\noindent with parameters \cite{Rin79}

\begin{eqnarray}
\hbar {\omega}_0 = 41 A^{-1/3} [\hbox{MeV}],
&\kappa_\pi =  0.0637, & \kappa_\nu = 0.0637, \\
&\mu_\pi = 0.60, &\mu_\nu = 0.42, \nonumber
\end{eqnarray}

The pairing interaction is

\begin{equation}
V_{pair} = -\frac{1}{4} G \sum_{j,j'} a^\dagger_j
a^\dagger_{\bar{j}} a_{\bar{j'}} a_{j'} \label{pair}
\end{equation}

\noindent where $\bar{j}$ denotes the time reversal partner of the
single-particle state $j$, and $G$ is the strength of the pairing
force. In principle we can start with an isospin invariant pairing
force, which contains proton-proton, neutron-neutron and
proton-neutron terms with equal strengths. However, in the limited
Hilbert space employed there are not protons and neutrons in the
same orbitals, and the proton-neutron pairing term is ineffective.
In recent works \cite{Var00a,Var00b,Var01,Pop00,Dra01}, the
pairing coefficients $G_{\pi,\nu}$ were fixed following
\cite{Rin79,Duf96}, with values:
\begin{equation}
G_\pi = \frac{21}{A} = 0.132 \hbox{~MeV}, ~~~~~~~~~
G_\nu = \frac{17}{A} = 0.106 \hbox{~MeV}.
\end{equation}

\section{The ground state of \gd ~ and \dy}

In this section we shall present the relevant results of the
pseudo SU(3) model, for the wave functions of the participant
nuclei. We shall assume, in subsection (III.A), that the wave
function describing the ground state of \gd ~ has only one
component and that the corresponding wave function for \dy ~has
two components connected by the pairing interaction. This
assumption allows us to perform a simple estimation of the double
beta decay, as shown in sections IV and V. In subsection (III.B)
we shall include other configurations in the model space, to asses
the quality of the simplified description.

\subsection{The one-component model}

 With a deformation of
$\epsilon = 0.26$ \cite{Moll95}, the most probable occupations for
the $^{160}$Gd 14 valence protons are 8 particles in normal and 6
particles in unique parity orbitals, and for the 14 valence
neutrons are 8 particles in normal and 6 particles in unique
parity orbitals.

After the detailed study of $^{156,158,160}$Gd isotopes performed in
\cite{Pop00,Dra01},
it becomes clear that the pseudo SU(3) model is a powerful tool in the description
of heavy deformed nuclei. Up to four rotational bands, with their intra- and inter-
band B(E2) transition strengths, as well as B(M1) transitions, were reproduced,
with a very good general agreement with the experiment.

The dominant component of the \gd ~ground state wave function
\cite{Pop00} is
\begin{eqnarray}
|^{160}\hbox{Gd}, 0^+\rangle  = &
 | \ (h_{11/2})^6_\pi,  \ J^A_\pi =  0; \
(i_{13/2})^6_\nu ,\ J^A_\nu =  0 \rangle_A \label{gd}\\
& | \{2^4\}_\pi (10,4)_\pi; \{2^4\}_\nu (18,4)_\nu; \ 1 (28,8) K=1, J = 0 \rangle_N.
\nonumber
\end{eqnarray}

In the first series of papers
\cite{Cas94,Hir95a,Hir95b,Hir95c,Hir95d} we have evaluated the
$\beta\beta$ matrix elements by taking into account only the
leading SU(3) coupled proton-neutron irrep. In recent publications
it was shown that the leading irrep represents approximately 60 \%
of the wave function in even-even Dy and Er isotopes
\cite{Pop00,Var01}.

Assuming a slightly larger deformation for $^{160}$Dy, the most probable occupations for
16 valence protons are 10 particles in normal and 6 particles in unique parity orbitals, and for the
12 valence neutrons are 6 particles in normal and 6 particles in unique parity orbitals.
A detailed study of $^{160,162,164}$Dy isotopes has been performed in \cite{Dra01}.

The dominant component of the wave function is
\begin{eqnarray*}
|^{160}\hbox{Dy}, 0^+ (a)\rangle  = &
 | \ (h_{11/2})^6_\pi,  \ J^A_\pi =  0; \ (i_{13/2})^6_\nu ,\ J^A_\nu =  0 \rangle_A \\
& | \{2^5\}_\pi (10,4)_\pi; \{2^3\}_\nu (18,0)_\nu; \ 1 (28,4) K=1, J = 0 \rangle_N.
\end{eqnarray*}
The two neutrino double beta operator annihilates two neutrons and
creates two protons with {\em the same} quantum numbers $\eta, l$.
It cannot connect the states $|^{160}\hbox{Gd},
 0^+\rangle$ and $|^{160}\hbox{Dy}, 0^+ (a)\rangle$ and the transition becomes
 absolutely forbidden. This is the selection rule found in \cite{Cas94}.

However, the pairing interaction allows for the mixing between
different occupations. In the deformed single particle Nilsson
scheme it takes an energy $\Delta E$ to promote a pair of protons
from the last occupied normal parity orbital to the next intruder
orbital. This excited state has 8 protons in normal and another  8
protons in unique parity orbitals, and its wave function has the
form
\begin{eqnarray*}
|^{160}\hbox{Dy}, 0^+ (b)\rangle  = &
 | \ (h_{11/2})^8_\pi,  \ J^A_\pi =  0; \ (i_{13/2})^6_\nu ,\ J^A_\nu =  0 \rangle_A \\
& | \{2^4\}_\pi (10,4)_\pi; \{2^3\}_\nu (18,0)_\nu; \ 1 (28,4) K=1, J = 0 \rangle_N.
\end{eqnarray*}
The two neutrino double beta decay of \gd ~can proceed to this
state.

As a first approximation, we shall describe the \dy  ~ ground
state as a linear combination of these two states:
\begin{equation}
|^{160}\hbox{Dy}, 0^+ \rangle =  a \,|^{160}\hbox{Dy}, 0^+ (a)\rangle  +
b \,|^{160}\hbox{Dy}, 0^+ (b)\rangle ,  \label{dy}
\end{equation}
with $|a|^2 + |b|^2 = 1$.

The only term in the Hamiltonian (\ref{hamil}) which can connect
states with different occupation numbers in the normal and unique
parity sectors is the pairing interaction. In the present case,
the Hamiltonian matrix has the simple form
\begin{equation}
H = \left(
\begin{array}{cc}  0  & h_{pair}  \\ h_{pair}  &\Delta E
\end{array}
 \right)
\label{hpair}
\end{equation}
with
\begin{eqnarray}
h_{pair} = \langle ^{160}\hbox{Dy}, 0^+ (b) | V_{pair} |^{160}\hbox{Dy}, 0^+ (a)\rangle
= {\frac {(-1)^{\tilde\eta_\pi +1} G_\pi}{4}}
\sqrt{(n^A_\pi + 2) (2 j^A_\pi + 1 - n^A_\pi)}
\sum_{\tilde l_\pi} \sqrt{2 (2 \tilde l_\pi + 1)}  \nonumber \\
\sum\limits_{(\lambda_0 \mu_0 )}
\langle (0 \tilde\eta_\pi )1 \tilde l_\pi,(0 \tilde\eta_\pi )1 \tilde l_\pi \|
(\lambda_0 \mu_0 ) 1 0 \rangle_1
\sum\limits_\rho \langle (\lambda^a, \mu^a) 1~0,(\lambda_0 \mu_0 )1~0 \|
(\lambda^b \mu^b )_\sigma 1 0 \rangle_\rho  \label{hmat}\\
\hspace{1cm}\sum\limits_{\rho_\pi}
\left[\begin{array}{cccc}
(\lambda^a_\pi, \mu^a_\pi) &(\lambda_0, \mu_0) &(\lambda^b_\pi, \mu^b_\pi) &\rho_\pi\\
(\lambda^a_\nu, \mu^a_\nu) &(0, 0 ) &(\lambda^b_\nu, \mu^b_\nu) & 1 \\
(\lambda^a, \mu^a) &(\lambda_0,\mu_0 ) &(\lambda^b \mu^b )_\sigma &\rho \\
1 &1 &1 \end{array} \right] \langle (\lambda^a_\pi,
\mu^a_\pi)\mid\mid\mid [\tilde a_{0,\tilde \eta_\pi),{1\over 2}}
\tilde a_{0,\tilde \eta_\pi),{1\over 2}}]^{(\lambda_0, \mu_0 )}
\mid\mid\mid (\lambda^b_\pi, \mu^b_\pi)\rangle_{\rho_\pi}
\nonumber
\end{eqnarray}

In the above formula $\langle ..,..\|,,\rangle$ denotes the SU(3)
Clebsch-Gordan coefficients \cite{Dra73}, the symbol  $[...]$
represents a $9-\lambda\mu$ recoupling coefficient \cite{Mil78},
and $\langle..\mid\mid\mid ..\mid\mid\mid ..\rangle$ is the triple
reduced matrix elements \cite{Hir95a}.

The lowest eigenstate has an energy
\begin{equation}
E = {\frac {\Delta E} {2}}
\left[ 1 - \sqrt{1 + \left( {\frac {2 h_{pair}} {\Delta E} } \right)^2} \right]~ñ,
\end{equation}
and the components of the \dy ~ground state wave function are
\begin{equation}
a = {\frac {h_{pair}}  { \sqrt{E^2 + h_{pair}^2} } }, ~~~
b = {\frac {E}  { \sqrt{E^2 + h_{pair}^2} } } ~.
\end{equation}

It is a limitation of the present model that $\Delta E$ has to be estimated
from the deformed Nilsson single particle mean field, instead of the
evaluation of the diagonal matrix element of the Hamiltonian (\ref{hamil}).
The use of seniority zero states for the
description of nucleons in intruder orbits inhibits the direct
comparison of states with different occupation numbers. A formalism
able to describe nucleons in both normal and unique parity orbitals
in the same footing is under development \cite{Var01b}. The effect
$\Delta E$ has on the double-beta-decay half-lives is studied in
Section VI.

\subsection{The enlarged model space}

In addition to the pseudo SU(3) irreps described in the previous
subsection, one has, in an enlarged model space, other potential
components to the ground state wave functions of \gd  ~and \dy .
In Tables I and II we are listing the four configurations included
in the description of \gd ~and the ones for \dy , respectively.
\begin{table}[h]
\begin{tabular}{c|cccc|cc|c|c}
\hline state&$n_{\pi}^N$& $n_{\pi}^A$& $n_{\nu}^N$&$n_{\nu}^A$
&($\lambda_{\pi}, \mu_{\pi}$)&($\lambda_{\nu},
\mu_{\nu}$)&($\lambda, \mu$)& C($\lambda,\mu$)\\
\hline $ i_1$& 8& 6& 8& 6& (10, 4)& (18,4)&(28,8)& 0.7253\\
$ i_2$ &8& 6&10& 4&(10, 4)& (20,4)&(30,8)&0.4848\\
$i_3$& 10& 4& 8&6& (10, 4)& (18,4)& (28,8)&
 0.4066\\
$i_4$& 10& 4& 10& 4& (10, 4)& (20,4)& (30,8)& 0.2713\\ \hline
\end{tabular}
\caption{The four configurations included in the description of
\gd . The number of proton
s and neutrons, in normal and in
intruder orbits and  the corresponding pseudo SU(3) irreps are
listed. The last column shows their amplitudes, in the wave
function of the ground state of \gd .}
\end{table}

The wave function, of the ground state of \gd , is written
\begin{equation}
\mid ^{160}Gd \rangle=\sum_k C_{k}^{(i)} \mid i_k \rangle ,
\end{equation}
and the one of \dy , is given by
\begin{equation}
\mid ^{160}Dy \rangle=\sum_k C_{k}^{(f)} \mid f_k \rangle ,
\end{equation}
where the amplitudes $C_{k}^{(i)}$ and $C_{k}^{(f)}$ are listed in
the last column of Tables I and II, respectively.
\begin{table}[h]
\begin{tabular}{c|cccc|cc|c|c}
\hline state&$n_{\pi}^N$& $n_{\pi}^A$& $n_{\nu}^N$&$n_{\nu}^A$
&($\lambda_{\pi}, \mu_{\pi}$)&($\lambda_{\nu},
\mu_{\nu}$)&($\lambda, \mu$)& C($\lambda,\mu$)\\
\hline $f_1$&10&6& 8& 4&(10, 4)& (18,4)& (28,8)& 0.7639\\
$f_2$& 10& 6& 6& 6& (10, 4)& (18,0)& (28,4)&0.5446\\
$f_3$& 8& 8& 8& 4& (10, 4)& (18,4)& (28,8)&0.2690\\
$f_4$ &8& 8&6& 6&(10, 4)& (18,0)&(28,4)&0.2181\\ \hline
\end{tabular}
\caption{ The four configurations included in the description of
\dy . The configurations and amplitudes are listed following the
notation given in the captions to Table I.}
\end{table}
The mixing matrix (\ref{hpair}), for the configurations of Tables
I and II, is a 4$\times$4 matrix, with matrix elements $h_{pair}$
of the form (\ref{hmat}). The diagonal matrix elements are the
energies needed to promote a pair of protons, neutrons or both
from a normal to an intruder parity orbital or viceversa. They are
estimated from the deformed single particle Nilsson diagrams. The
values of $\Delta E$, for the case of \gd ~are 0.54 MeV, 0.81 MeV
and 1.35 MeV. The corresponding quantities for \dy ~are of the
order of 0.54 MeV, 1.71 MeV and 2.25 MeV. The diagonalization of
the mixing matrix yields the amplitudes $C$ of the wave functions,
and they are shown in the last column of Tables I and II. As one
can see from these tables, the mixing induced by the pairing
interaction is sizeable although the dominant component of each
wave function is still the irrep considered in the restricted
model space.

Naturally, and in order to estimate the effect of this mixing upon
the double beta decay process, we should compute explicitly the
corresponding matrix elements, which are the sum of products of
amplitudes and individual matrix elements. It will be done in the
following sections.

\section{ The \bt ~half-life}

The inverse half-life of the two neutrino mode of the
$\beta\beta$-decay, \bt , can be cast in the form \cite{Doi85}

\begin{equation}
    \left[\tau^{1/2}_{2\nu}(0^+ \rightarrow 0^+)\right]^{-1} =
      G_{2\nu} \ | \ M_{2\nu} \ |^2 \ \ .
\end{equation}

\noindent where $G_{2\nu}$ is a kinematical factor which depends
on ${\it Q}_{\beta  \beta}$, the total kinetic energy released in
the decay.

The nuclear matrix element is written
\begin{equation}
M_{2\nu} \approx M_{2\nu}^{GT} = \sum_{N}
{ { \langle  0^+_f \ || \, \Gamma \, || \ 1^+_N \rangle \  \langle 1^+_N \
 || \,\Gamma \, ||\  0^+_i \rangle \,} \over{\mu_N} } , \label{m1}
\end{equation}
with the Gamow-Teller operator $ \Gamma$ expressed as
\begin{equation}
 \Gamma_m = \sum_s \ \sigma_{ms} t^-_s \equiv \sum_{\pi \, \nu}
 \sigma (\pi ,\nu ) [a^{\dagger}_{\eta_{\pi} l_{\pi} \, {1 \over 2};j_{\pi}}
\otimes \tilde a_{\eta_{\nu} l_{\nu} \, {1 \over 2};j_{\nu}}]^1_m
, \hspace{1cm}m=1,0,-1.   \label{gt}
\end{equation}
  The  energy denominator is $\mu_N = E_f + E_N -E_i$  and it
contains the intermediate $E_N$, initial $E_i$ and final $E_f$
energies. The kets $|1^+_N\rangle$  denote intermediate states.

The mathematical expressions needed to evaluate the nuclear matrix
elements of the allowed $g.s. \rightarrow g.s.$ \bb ~decay in the
pseudo SU(3) model were developed in \cite{Cas94}. Using the
summation method described in \cite{Cas94,Civ93}, exploiting the
fact that the two body terms of the  $\tilde{SU(3)}$ Hamiltonian
commute with the Gamow-Teller operator (\ref{gt}) \cite{Hir95a},
resumming the infinite series and recoupling the Gamow-Teller
operators, the following expression was found:
\begin{equation}
\begin{array}{ll}
M_{2\nu}^{GT} =
& \sqrt{3}  \sum\limits_{\pi\nu ,\pi' \nu'}
{{\sigma (\pi ,\nu )\sigma (\pi',\nu')}
\over {(E_0 +\epsilon_\pi -\epsilon_\nu)} }
 \langle  0^+_f |
\left[ [a^{\dagger}_\pi \otimes \tilde a_\nu ]^1 \otimes
[a^{\dagger}_{\pi'} \otimes \tilde a_{\nu'}]^1 \right]^{J=0}
|\  0^+_i \rangle
\label{m2} \end{array}
\end{equation}

\noindent
where $\pi \equiv (\eta_{\pi},l_{\pi},j_{\pi})$ and $\nu \equiv
(\eta_{\nu},l_{\nu},j_{\nu})$, and
$E_0 = {\frac {{\it Q}_{\beta  \beta}} {2}} + m_e c^2$ .

In the following we shall analyze the nuclear matrix element
(\ref{m2}) for the \bt ~decay of the ground state of \gd , Eq.
(\ref{gd}), to the ground state of \dy , Eq.(\ref{dy}). Each
Gamow-Teller operator (\ref{gt}) annihilates a proton and creates
a neutron in the same oscillator shell and with the same orbital
angular momentum. In the case of the \bb ~of \gd ~ it means that
the operator annihilates two neutrons in the pseudo shell
$\eta_\nu = 5$ and creates two protons in the abnormal orbit
$h_{11/2}$. As a consequence the only orbitals which in the model
space can be connected by the \bb ~decay are those satisfying $
\eta_{\pi} = \eta_{\nu} \equiv \eta$, that implies $l_\pi = l_\nu
= \eta$, $j_\nu = \eta - {1 \over 2}$ and $j_\pi = \eta + {1 \over
2}$.

Under these restrictions the \bt ~decay is allowed only if the
occupation numbers obey the following relationships
\begin{equation}
\begin{array}{l}
n^A_{\pi ,f} = n^A_{\pi ,i} + 2~~,
\hspace{1cm}n^A_{\nu ,f} = n^A_{\nu ,i}~~, \\
n^N_{\pi ,f} = n^N_{\pi ,i}~~ ,
\hspace{1.7cm} n^N_{\nu ,f} = n^N_{\nu ,i} - 2 ~~.\label{num}
\end{array}
\end{equation}
It follows that, when using the restricted configuration space in
Eq. (\ref{m2}), only one term in the sum survives and thus the
nuclear matrix element $M_{2 \nu}$ (\ref{m2}) can be written as
\begin{equation}
M_{2\nu}^{GT} = \frac {{\sigma(\pi,\nu)}^2} {\cal E}  \langle
0^+_f | \left[ [a^{\dagger}_\pi \otimes \tilde a_\nu ]^1 \otimes
[a^{\dagger}_{\pi} \otimes \tilde a_{\nu}]^1 \right]^{J=0} | 0^+_i
\rangle ,\label{m3}
\end{equation}
where the energy denominator ${\cal E}$ is determined by demanding
that the energy of the Isobaric Analog State equals the difference
in Coulomb energies $\Delta_C$. It is given by \cite{Cas94}
\begin{equation}
\begin{array}{ll}
{\cal E} = &E_0 + \epsilon (\eta_\pi,l_\pi ,j_\pi = j_\nu + 1) -
\epsilon (\eta_\nu ,l_\nu ,j_\nu )=E_0
 -\hbar \omega k_\pi 2 j_\pi + \Delta_C .\\
~\\
&\Delta_C ={ 0.70 \over A^{1/3}} [2 Z + 1 - 0.76 ( (Z+1)^{4/3} -Z^{4/3} )]
  MeV. \label{den}
\end{array}
\end{equation}

As it was discussed in \cite{Cas94}, Eq. (\ref{m3}) has no free
parameters, being the denominator (\ref{den}) a well defined
quantity. The reduction to only one term is a consequence of the
restricted Hilbert proton and neutron spaces of the model. The
initial and final ground states are strongly correlated with  a
very rich structure in terms of their shell model components.

The low energy levels are assumed to have pseudo spin $\tilde S =
0$, a fact that again simplifies the evaluation of the above sum
since $\tilde L = J$.

The calculation of the  matrix elements, Eq. (\ref{m3}), in the
normal space is performed by using $SU(3)$ Racah calculus to
decouple the proton and neutron normal irreps, and expanding the
annihilation operators in their SU(3) tensorial components. For
the particular case of \gd , in the spirit of the single model
space of subsection III.A, the \bt ~decay can reach only  the
second component of the wave function of the ground state of \dy ~
(\ref{dy}), and for this reason it is proportional to the
amplitude $b$. The expression for the matrix elements of the \bt ~
channel reads
\begin{equation}
\begin{array}{ll}
M_{2\nu}^{GT}($\gd $\rightarrow $\dy$)~ =~b ~{\frac{2 j_\pi -1}{2 j_\pi}}
\left[{\frac{(n^A_\pi + 2)(2 j_\pi +1 - n^A_\pi)}{ j_\pi -1}}\right]^{1/2}
~{\cal E}^{-1}\\
\hspace{1cm}\sum\limits_{(\lambda_0, \mu_0 )}
\langle (0 \tilde\eta )1 \tilde l,(0 \tilde\eta )1 \tilde l \| (\lambda_0 \mu_0
) 1 0\rangle_1 \sum\limits_\rho
\langle (28,8)1~0,(\lambda_0,\mu_0 )1~0\|(28, 4 ) 1 0\rangle_\rho\\
\hspace{1cm}\sum\limits_{\rho'}
\left[\begin{array}{cccc} (10,4) &(0,0) &(10,4) &1\\
(18,4) &(\lambda_0 \mu_0 ) &(18,0) &\rho' \\
(28,8) &(\lambda_0 \mu_0 ) &(28,4 ) &\rho \\
1 &1 &1 \end{array} \right] \langle (18,0)\mid\mid\mid [\tilde
a_{0 \tilde \eta),{1\over 2}} \tilde a_{0 \tilde \eta),{1\over
2}}]^{(\lambda_0 \mu_0 )} \mid\mid\mid (18,4)\rangle_{\rho'}  .
\label{mgt}
\end{array}
\end{equation}

In the so-called large model space, of subsection III.B, the
expression for the matrix element of the \bt ~decay mode is

\begin{equation}
\begin{array}{ll}
M_{2\nu}^{GT}($\gd $\rightarrow $\dy$)~ =~ \sum\limits_{k,l}
C_k^{(i)}C_l^{(l)} ~{\frac{2 j_{\pi,k} -1}{2 j_{\pi,k}}}
\left[{\frac{(n^A_{\pi,k} + 2)(2 j_{\pi,k} +1 - n^A_{\pi,k})}{
j_{\pi,k} -1}}\right]^{1/2}
~{\cal E}^{-1}\\
\hspace{1cm}\sum\limits_{(\lambda_0, \mu_0 )} \langle (0
\tilde\eta )1 \tilde l,(0 \tilde\eta )1 \tilde l \| (\lambda_0
\mu_0 ) 1 0\rangle_1 \sum\limits_\rho
\langle (\lambda_k,\mu_k)1~0,(\lambda_0,\mu_0 )1~0\|(\lambda_l, \mu_l ) 1 0\rangle_\rho\\
\hspace{1cm}\sum\limits_{\rho'}
\left[\begin{array}{cccc} (\lambda_{\pi,k},\mu_{\pi,k}) &(0,0)
&(\lambda_{\pi,l},\mu_{\pi,l}) &1\\
(\lambda_{\nu,k},\mu_{\nu,k}) &(\lambda_0 \mu_0 ) &(\lambda_{\nu,l},\mu_{\nu,l}) &\rho' \\
(\lambda_k,\mu_k) &(\lambda_0 \mu_0 ) &(\lambda_l,\mu_l) &\rho \\
1 &1 &1 \end{array} \right] \langle (\lambda_l,\mu_l)\mid\mid\mid
[\tilde a_{0 \tilde \eta),{1\over 2}} \tilde a_{0 \tilde
\eta),{1\over 2}}]^{(\lambda_0 \mu_0 )} \mid\mid\mid
(\lambda_k,\mu_k)\rangle_{\rho'} . \label{mgtl}
\end{array}
\end{equation}

\section{ The \bz half-life}

For massive Majorana neutrinos one can perform the integration
over the four-momentum of the exchanged particle and obtain a
``neutrino potential'' which for a light neutrino ($m_\nu < 10$
MeV) has the form

\begin{equation}
H(r,\overline E) = {\frac {2R}{\pi r}} \int_0^\infty dq {\frac {sin(qr)}
{q+\overline E} }  ~,
\end{equation}

\noindent where $\overline E$ is the average excitation energy of
the intermediate odd-odd nucleus and the nuclear radius $R$ has
been added to make the neutrino potential dimensionless. In the
zero neutrino case this closure approximation is well justified
\cite{Pan92}. The final formula, restricted to the term
proportional to the neutrino mass, is \cite{Ver86,Doi85}
\begin{equation}
(\tau^{1/2}_{0\nu})^{-1} = \left ( {\frac {\langle m_\nu \rangle}{m_e}} \right )^2
G_{0\nu}  M_{0\nu}^2   ~.
\end{equation}

\noindent
where $G_{0\nu}$ is the phase space integral associated with the emission
of the two electrons. The nuclear matrix elements
$M_{0\nu}$ are \cite{Doi85}

\begin{equation}
M_{0\nu} \equiv | M_{0\nu}^{GT} - {\frac{g_V^2}{g_A^2}} M_{0\nu}^{F} |  ~, \label{m0nu}
\end{equation}
\noindent
with

\begin{equation}
M_{0\nu}^\alpha =  \langle  0^+_f \| O^\alpha \| 0^+_i\rangle  ~,
\end{equation}
\noindent where the kets
$ \vert 0^+_i \rangle$  and $|0^+_f\rangle$ denote
the corresponding initial and final nuclear  states, the quantities $g_V$
and $g_A$ are the dimensionless coupling constants of the vector and
axial vector nuclear currents, and

\begin{equation}
\begin{array}{l}
O^{GT} \equiv \sum\limits_{m,n} O^{GT}_{mn} = \sum\limits_{m,n}
\vec\sigma_m
t^-_m \cdot \vec\sigma_n t^-_n H(|\vec r_m - \vec r_n |,\overline E) ~,\\
O^{F} \equiv \sum\limits_{m,n} O^{F}_{mn} = \sum\limits_{m,n}  t^-_m t^-_n
H(|\vec r_m - \vec r_n |,\overline E)  ~,
\end{array}
\end{equation}

\noindent being $\vec\sigma$ the Pauli matrices related with the
spin operator and $t^-$ the isospin lowering operator, which
satisfies $t^-|n\rangle = |p\rangle$. The superindex GT denotes
the Gamow-Teller operator, while F indicates the Fermi operator.
In the present work we use the effective value $({\frac {g_A}
{g_V}})^2 = 1.0$ \cite{Vog86}.

Transforming the transition operators to the pseudo $SU(3)$ space,
we have the formal expression
\begin{equation}
O^\alpha =  O^\alpha_{N_\pi N_\nu} +O^\alpha_{N_\pi A_\nu} +
O^\alpha_{A_\pi N_\nu} +O^\alpha_{A_\pi A_\nu}
\end{equation}
\noindent where the subscript index $NN, \ NA, \dots$ are
indicating the normal or abnormal spaces of the nucleon creation
and annihilation operators, respectively. Given that we use the
Nilsson scheme to obtain the occupation numbers, we are
considering only nucleon pairs. For this reason only the four type
of transitions listed above give a non-vanishing contribution to
the \bt ~matrix elements.

In a previous work we have restricted our analysis to six
potential double beta emitters which, within the approximations of
the simplest pseudo SU(3) scheme, are also decaying via the $2\nu$
\bb ~ mode. They include the observed $^{150}$Nd $ \rightarrow$ $
^{150}$Sm and $^{238}$U $\rightarrow$ $ ^{238}$Pu \bt ~decays. In
this case two neutrons belonging to a normal parity orbital decay
in two protons belonging to an abnormal parity orbital. The
transition is mediated by the operator $O^\alpha_{A_\pi N_\nu}$.
Under the seniority zero assumption for nucleons in abnormal
parity orbitals, only proton pairs coupled to J=0 are allowed in
the ground state. The matrix elements is
\begin{eqnarray}
M_{0\nu}^\alpha(A_\pi N_\nu) &\equiv \sum\limits_{j_\nu} M_{0\nu}^\alpha (A_\pi N_\nu,j_\nu ) =
 -\sum\limits_{j_\nu} \sqrt{\frac {2 j_\nu +1} {2(2\tilde l_\nu + 1)}}
\sqrt{{\frac {(n^A_\pi+2)(2j_\pi +1-n^A_\pi)} {2 j_\pi + 1}}}
\langle (j^A_\pi j^A_\pi )J=M=0 |O^\alpha |(j_\nu j_\nu ) J=M=0\rangle  \nonumber \\
&\sum\limits_{K_\pi L_\pi} \sum\limits_{K_\nu K'_\nu}
\langle (\lambda_\pi^f,\mu_\pi^f)K_\pi L_\pi ,
(\lambda_\nu^f,\mu_\nu^f)K'_\nu L_\pi \|(\lambda^f,\mu^f) 1 0 \rangle
\langle (\lambda_\pi^i,\mu_\pi^i)K_\pi L_\pi ,
(\lambda_\nu^i,\mu_\nu^i)K_\nu L_\pi \|(\lambda^i,\mu^i) 1 0 \rangle  \nonumber \\
&\sum\limits_{(\lambda_0 ,\mu_0 ) \rho_0}
\langle (\lambda_\nu^i,\mu_\nu^i)K_\nu L_\nu , (\lambda_0 ,\mu_0 ) 1 0 \|
(\lambda_\nu^f,\mu_\nu^f) K'_\nu L_\nu \rangle_{\rho_0}
\langle  (0,\tilde \eta_\nu) 1 \tilde l_\nu , (0,\tilde \eta_\nu) 1
\tilde l_\nu \| (\lambda_0 ,\mu_0 ) 1 0\rangle \label{m0nu-me} \\
& \langle (\lambda_\nu^f,\mu_\nu^f) \|| [\tilde a_{(0,\tilde
\eta_\nu) \tilde l_\nu ; {\frac 1 2}} \tilde a_{(0,\tilde
\eta_\nu) \tilde l_\nu ; {\frac 1 2}}]^{ (\lambda_0 ,\mu_0
);\tilde S = 0} \|| (\lambda_\nu^i,\mu_\nu^i)\rangle_{\rho_0}  .
\nonumber
\end{eqnarray}
 We have implicitly defined
$M_{0\nu}^\alpha (j_\nu) $ as the contribution of each normal
parity neutron state $j_\nu$ to the nuclear matrix element in the transition
$(j_\nu )^2 \rightarrow (j_\pi^A)^2$.

The transitions which are forbidden for the \bt ~decay are allowed
for the zero neutrino mode, due to presence of the neutrino
potential, instead. In the simplest model space, the \bz ~of \gd ~
has finite contributions for the two components with different
occupation numbers in the \dy  ~ final state. There are two terms
in the \bz ~decay: one to the basis state which has allowed \bt
~decay, and one to the state with forbidden \bt ~decay. In the
first case the above equation must be used. The second case
involves the annihilation of two neutrons in normal parity
orbitals, and the creation of two protons in normal parity
orbitals. This transition is mediated by the operator
$O^\alpha_{N_\pi N_\nu}$. The \bz ~matrix element has the form
\begin{eqnarray}
M_{0\nu}^\alpha (N_\pi N_\nu) &\equiv \sum\limits_{J} M_{0\nu}^\alpha (N_\pi N_\nu,J )
= - {\frac 1 4}
 \sum\limits_{j_\pi j_{\pi'} j_\nu j_{\nu'}}
\sqrt{(2 j_{\pi} +1) (2 j_{\pi'} +1) (2 j_{\nu} +1)(2 j_{\nu'} +1)} \nonumber \\
& \sum\limits_J \sqrt{2 J + 1} ~
\langle (j_\pi j_{\pi'} )J |O^\alpha |(j_\nu j_{\nu'} ) J\rangle  ~
W(\tilde l_\pi J {\frac 1 2} j_{\pi'}, \tilde l_{\pi'} j_{\pi}) ~
W(\tilde l_\nu J {\frac 1 2} j_{\nu'}, \tilde l_{\nu'} j_{\nu})   \nonumber \\
&\sum\limits_{(\lambda_0^\pi ,\mu_0^\pi ) K_0^\pi}
\langle  (\tilde \eta_\pi, 0) 1 \tilde l_\pi , (\tilde \eta_\pi , 0) 1
\tilde l_{\pi'} \| (\lambda_0^\pi ,\mu_0^\pi ) K_0^\pi J \rangle
\sum\limits_{(\lambda_0^\nu ,\mu_0^\nu ) K_0^\nu}
\langle  (0,\tilde \eta_\nu) 1 \tilde l_\nu , (0,\tilde \eta_\nu) 1
\tilde l_{\nu'} \| (\lambda_0^\nu ,\mu_0^\nu ) K_0^\nu J \rangle  \\
&\sum\limits_{(\lambda_0 ,\mu_0 ) \rho_0}
\langle (\lambda_0^\pi ,\mu_0^\pi ) K_0^\pi J , (\lambda_0^\nu ,\mu_0^\nu ) K_0^\nu J \|
(\lambda_0 ,\mu_0 ) 1 0  \rangle_{\rho_0}
\sum\limits_{\rho}
\langle (\lambda^i,\mu^i) 1 0, (\lambda_0 ,\mu_0 ) 1 0 \|
(\lambda^f,\mu^f) 1 0 \rangle_{\rho} \nonumber \\
& \sum\limits_{\rho_\pi \rho_\nu}
\left[\begin{array}{cccc}
(\lambda^i_\pi, \mu^i_\pi) &(\lambda_0^\pi , \mu_0^\pi) &(\lambda^f_\pi, \mu^f_\pi) &\rho_\pi\\
(\lambda^i_\nu, \mu^i_\nu) &(\lambda_0^\nu ,\mu_0^\nu ) &(\lambda^f_\nu, \mu^f_\nu) & \rho_\nu \\
(\lambda^i, \mu^i) &(\lambda_0,\mu_0 ) &(\lambda^f \mu^f ) &\rho \\
1 &\rho_0 &1 \end{array} \right]
\langle (\lambda^f_\pi, \mu^f_\pi)\mid\mid\mid [a^\dagger_{\tilde \eta_\pi , 0),{1\over 2}}
a^\dagger_{\tilde \eta_\pi , 0),{1\over 2}}]^{(\lambda_0^\pi ,\mu_0^\pi ) ;\tilde S =0}
\mid\mid\mid (\lambda^i_\pi, \mu^i_\pi)\rangle_{\rho_\pi} \nonumber \\
&\langle (\lambda_\nu^f,\mu_\nu^f) \||
[\tilde a_{(0,\tilde \eta_\nu) \tilde l_\nu ; {\frac 1 2}}
\tilde a_{(0,\tilde \eta_\nu) \tilde l_\nu ; {\frac 1 2}}]^{
(\lambda_0^\nu ,\mu_0^\nu );\tilde S = 0} \||
(\lambda_\nu^i,\mu_\nu^i)\rangle_{\rho_\nu}   ~.  \nonumber
\end{eqnarray}
In the above expression the $W(...,...)$ are Racah Coefficients \cite{Var88}.
The two-body matrix element can be expanded in its $L,S$ components
\begin{eqnarray}
\langle (j_\pi j_{\pi'} )J |O^\alpha |(j_\nu j_{\nu'} ) J \rangle
=   \nonumber\\
\sum_{ L S}
\chi\left\{ \begin{array}{ccc}
l_\pi & l_{\pi'} & L \\ {\frac 1 2} &{\frac 1 2} &S \\ j_\pi &j_{\pi'} & J
\end{array} \right\}
\chi\left\{ \begin{array}{ccc}
l_\nu & l_{\nu'} & L \\ {\frac 1 2} &{\frac 1 2} &S \\ j_\nu &j_{\nu'} & J
\end{array} \right\} ~
\langle (l_\pi l_\pi )L \| H(r, \overline E) \| (l_\nu l_\nu) L \rangle ~
\langle ({\frac 1 2}{\frac 1 2}) S \| \Gamma \cdot \Gamma (\alpha) \|
({\frac 1 2}{\frac 1 2}) S \rangle
\end{eqnarray}
where the $\chi\left\{...\right\}$ are  Jahn-Hope coefficients \cite{Var88} and
\begin{equation}
\Gamma \cdot \Gamma (GT) \equiv \vec\sigma_1 \cdot \vec\sigma_2
\hspace{1cm}
\Gamma \cdot \Gamma (F) \equiv 1
\hspace{1cm}
\alpha = GT ~\hbox{or}~ F   .
\end{equation}

In order to evaluate the spatial matrix elements, we have used the
Bessel-Fourier expansion of the potential \cite{Hor61}, which
gives
\begin{eqnarray}
\langle (l_1 l_2 ) L M | H(r) | (l_3 l_4 ) L M \rangle
= \sum_l (-1)^{l_1 + l_4 + L} ~(2l+1) ~(l_1\|C_l\|l_3)~ (l_2\|C_l\|l_4)~
W(l_1 l_2 l_3 l_4;L l)~  R^l(l_1 l_2,l_3l_4)
\end{eqnarray}
where $(l_i\|C_l\|l_j)$ are the reduced matrix elements of the unnormalized
spherical harmonics
$ C_{lm}(\Omega ) \equiv \sqrt{\frac {4\pi}{2l+1}} Y_{lm}(\Omega )$
and $R^l(l_1 l_2 , l_3 l_4)$ are the radial integrals described in
Appendix A of Ref. \cite{Hir95a}. They include the effects of the finite
nucleon size and the short range correlations, as explained in Appendix C
of Ref. \cite{Hir95a}.

The zero neutrino \bb ~matrix element has, as mentioned above,
contributions from the two components of the \dy  ~ wave function:
\begin{equation}
M_{0\nu}^\alpha ~ = ~a ~M_{0\nu}^\alpha (N_\pi N_\nu) ~+ ~b ~
M_{0\nu}^\alpha (A_\pi N_\nu) \label{monu}
\end{equation}

As said before, these expressions are to be supplemented by other
contributions, $M_{0\nu}^{\alpha}(N_\pi A_\nu)$ and
$M_{0\nu}^{\alpha}(A_\pi A_\nu)$, when the large model space (of
subsection III.B) is used to construct the initial and final wave
functions. The expression of the additional terms is similar to
the ones of the above equations and they are omitted for the sake
of brevity. The matrix element of the \bz ~mode is thus given by

\begin{equation}
M_{0\nu}^\alpha ~ = \sum_{k,l}C_k^{(i)}C_l^{(f)} ~M_{0\nu}^\alpha
(k, l) ,
\end{equation}
where the indexes $k$ and $l$ denote the set of quantum numbers
needed to specify the occupations and irreps included in the wave
functions.

\section{The \bb ~of \gd }

In this section we study the two neutrino and zero neutrino modes
of the double beta decay of \gd ~to the ground state of \dy .

In the restricted configuration space of subsection III.A, the
ground state of \dy , defined in Eq.(\ref{dy}), is a linear
combination of two states having different occupation numbers. The
energy difference between these states, estimated from the
difference in their deformed Nilsson single particle energies, is
$\Delta E = 1.71$ MeV. The pairing mixing between them, using the
interaction strength $G_\pi = 21 / A $ MeV, is $h_{pair} = 0.865$
MeV. With these matrix elements, the diagonalization of Eq. (9)
yields the amplitudes $a = 0.923, ~ b= 0.385$ of the wave function
of the ground state of \dy .

The \bt ~ matrix element is suppressed by a factor $b$, compared
with the {\em allowed} decays. It implies that the \bt ~half-life
is an order of magnitude ($1/b^2$) larger than in other nuclei
with similar $Q_{\beta\beta}$ values. The energy denominator takes
the value ${\cal E} = 12.19$ MeV. The two neutrino \bb ~matrix
element is $M^{GT}_{2\nu}$(\gd $\rightarrow$ \dy) = 0.0455
MeV$^{-1}$. The phase space integral is $G_{GT} = 8.028 \times
10^{-20}$ MeV$^2$ yr$^{-1}$, using $g_A/g_V = 1.0$. The estimated
\bt ~half-life is
\begin{equation}
\tau^{1/2}_{2\nu} ( \hbox{\gd} \rightarrow \hbox{\dy}) = 6.02 \times 10^{21} \hbox{yr}.
\end{equation}

The contribution of the different angular momentum $J$ to the
 matrix elements $M_{0\nu}^\alpha (N_\pi N_\nu,J )$, in Eq.
( \ref{monu}), are shown in Table III.

\begin{table}
\begin{tabular}{crr}
    J     &   ~~~~~~~~ F ~~~   &  ~~~~~~~~~ GT ~~ \\
\hline
    0   &  -0.11746  &  0.25588  \\
    2   &   0.03792  & -0.05016  \\
    3   &  -0.00131  & -0.00011  \\
    4   &   0.02090  &  0.00454  \\
    6   &   0.00312  &  0.00501  \\  %\\
Sum     &  -0.05683  &  0.21516
\end{tabular} \\  ~~\\
$ M_{0\nu} (N_\pi N_\nu)$ =    0.27199
\caption{$M_{0\nu}^\alpha (N_\pi N_\nu,J )$ for the \bz ~of \gd .}
\end{table}
It is remarkable that the J=0 channel largely dominates both the
Fermi (F) and Gamow-Teller (GT) transitions. The J = 2
channel tends to reduce the transition matrix elements by about
20 to 30 \%, For the Fermi matrix elements, the J = 4 channel also
contributes noticeably. Fermi and Gamow-Teller matrix elements
add coherently due to the sign inversion in Eq. (\ref{m0nu}).

The transition matrix elements from the different neutron single
particle angular momentum $j_\nu$ in the matrix elements
$M_{0\nu}^\alpha (A_\pi N_\nu, j_\nu )$, are presented in Table
IV. In this case two neutrons are annihilated in the normal parity
orbitals $p_{1/2}, p_{3/2}, f_{5/2}, f_{7/2}, h_{9/2}$ and two
protons are created in the intruder orbit $h_{11/2}$.

\begin{table}
\begin{tabular}{crr}
 $~~~j_\nu$ ~~~  &  ~~~~~~~~    F ~~~  &  ~~~~~~~~~  GT ~~ \\
\hline
 $p_{1/2}$   &  0.01552 & -0.06792  \\
 $p_{3/2}$   &  0.02105 & -0.05359  \\
 $f_{5/2}$   &  0.02168 & -0.16133  \\
 $f_{7/2}$   &  0.06341 & -0.11847  \\
 $h_{9/2}$   &  0.04310 & -1.16598  \\
Sum          &  0.16477 & -1.56729
\end{tabular} \\  ~~\\
$ M_{0\nu} (A_\pi N_\nu)$ =   -1.73206
\caption{$M_{0\nu}^\alpha (A_\pi N_\nu,j_\nu )$ for the \bz ~of \gd .}
\end{table}
As seen in Table IV, both for the Fermi and Gamow-Teller matrix
elements the terms add coherently. While in the Gamow-Teller case
the $h_{9/2} \rightarrow h_{11/2}$ transition clearly dominates,
in the Fermi case all transitions amplitudes are comparable. As in
Table III,  Fermi and Gamow-Teller final matrix elements add
coherently. In absolute value, the {\em allowed} \bz ~transition
matrix element $ M_{0\nu} (A_\pi N_\nu)$ is a factor 6 larger than
the {\em forbidden} one $ M_{0\nu} (N_\pi N_\nu)$. The final \bz
~matrix element is
\begin{equation}
M_{0\nu} ( \hbox{\gd} \rightarrow \hbox{\dy}) = a \, M_{0\nu}
(N_\pi N_\nu) + b\,  M_{0\nu} (A_\pi N_\nu) = 0.251 + 0.668 =
0.919 .
\end{equation}
The zero neutrino phase space integral for the \bz ~ of \gd ~is
$G_{0\nu} = 1.480 \times 10^{-14}$ yr$^{-1}$. Expressing the
Majorana mass of the neutrino $\langle m_\nu \rangle$ in units of
eV , the calculated zero neutrino \bb ~half-life is
\begin{equation}
\tau^{1/2}_{0\nu} ( \hbox{\gd} \rightarrow \hbox{\dy}) \cdot \langle m_\nu \rangle^2
 = 2.09 \times 10^{25} \hbox{yr}.
\end{equation}
This half-life is a factor 20 larger than the one reported in
\cite{Sta90}. The difference is easily understood, given the fact
that the present model takes explicitly into account the nuclear
deformation, while in \cite{Sta90} the spherical QRPA was used to
obtain a crude estimations of the half-lives of all potential \bb
~ emitters.

As mentioned above, the parameter $\Delta E$, which strongly
influences the pairing mixing, is taken from the deformed Nilsson
single particle energies. To estimate in which extent changes in
this parameter affect the predicted double-beta-decay half-lives,
we have allowed it to vary from 1.1 MeV to 2.5 MeV, covering most
of the physically reasonable range. The results are listed in
Table V.

\begin{table}
\begin{tabular}{cc|cc|cc}
$\Delta E$[MeV]   &~~$b$~~~~  & $M^{GT}_{2\nu}$[MeV$^{-1}$]  &~~$\tau^{1/2}_{2\nu}$[10$^{21}$ yr]
 & ~~$M_{0\nu}$~~  &$\tau^{1/2}_{0\nu}$[10$^{25}$ yr] \\ \hline
1.10    &0.481      &0.0568     &~3.86      &1.072      &1.53  \\
1.20    &0.464      &0.0547     &~4.16      &1.044      &1.62  \\
1.30    &0.447      &0.0527     &~4.48      &1.017      &1.70  \\
1.40    &0.431      &0.0508     &~4.82      &0.992      &1.79  \\
1.50    &0.415      &0.0490     &~5.18      &0.967      &1.89  \\
1.60    &0.401      &0.0473     &~5.57      &0.943      &1.98  \\
1.70    &0.387      &0.0456     &~5.98      &0.921      &2.08  \\ \hline
1.71    &0.385      &0.0455     &~6.02      &0.919      &2.09  \\ \hline
1.80    &0.374      &0.0441     &~6.41      &0.900      &2.18  \\
1.90    &0.361      &0.0426     &~6.86      &0.879      &2.28  \\
2.00    &0.349      &0.0412     &~7.34      &0.860      &2.38   \\
2.10    &0.338      &0.0399     &~7.83      &0.841      &2.49  \\
2.20    &0.327      &0.0386     &~8.36      &0.824      &2.60  \\
2.30    &0.317      &0.0374     &~8.90      &0.807      &2.71  \\
2.40    &0.307      &0.0363     &~9.47      &0.791      &2.82  \\
2.50    &0.298      &0.0352    &10.06       &0.776      &2.93
\end{tabular} \\  ~~\\
\caption{The mixing parameter $b$, the double-beta-decay matrix elements and
half-lives are listed as functions of the parameter $\Delta E$.}
\end{table}
The mixing parameter $b$ and the double-beta-decay matrix elements
and half-lives are very smooth functions of $\Delta E$. The two
neutrino double-beta-decay half-life varies between 4 $\times
10^{21}$ yr and 10 $ \times 10^{21}$ yr, around the predicted
value of 6 $ \times 10^{21}$ yr. The calculated half-life of the
neutrinoless double-beta-decay is less dependent upon the mixing
induced by pairing and it varies in the range  1.5 - 2.9 $ \times
10^{25}$ yr.

Concerning the results obtained in the enlarged space of
subsection III.B, we are presenting them in Table VI. The partial
contributions to the \bt ~and \bz ~ decay processes, for
transitions between the components of the initial and final wave
functions, are listed in this table. We are indicating, also, the
character of each of the transitions, with regard to the type of
orbital, abnormal (A) or normal (N), of the neutrons and protons
involved in the decay.

\begin{table}
\begin{tabular}{cc|cc}
\hline Transition& channel& 2$\nu \beta \beta$ mode & 0 $\nu \beta
\beta$\\
\hline
$i_1 \rightarrow f_1$&$N_{\pi} A_{\nu}$& not allowed &-0.231\\
$i_1 \rightarrow f_2$ &$N_{\pi} N_{\nu}$&not allowed& 0.272\\
$i_1 \rightarrow f_3$& $A_{\pi} A_{\nu}$&not allowed&-2.122\\
$i_1 \rightarrow f_4$& $A_{\pi} N_{\nu}$& 0.118 & 1.732\\
$i_2 \rightarrow f_1$&$N_{\pi} N_{\nu}$&not allowed& 0.315\\
$i_2 \rightarrow f_3$& $A_{\pi} N_{\nu}$&0.122&1.787\\
$i_3 \rightarrow f_1$& $A_{\pi} A_{\nu}$&not allowed &-2.122\\
$i_3 \rightarrow f_2$& $A_{\pi}N_{\nu}$&0.118&1.732\\
$i_4 \rightarrow f_1$& $A_{\pi} N_{\nu}$&0.122&1.787\\
\hline
\end{tabular}
\caption{Matrix elements for the \bt ~and \bz ~decay modes. Each
arrow shows the participant configurations, of the initial and
final wave functions, which are listed in Tables I and II. The
transitions not included in the table are trivially forbidden
because they imply the change of the states of more than two
nucleons. The \bt ~transitions listed as ``not allowed '', are the
ones which are forbidden by the selection rules discussed in
section IV. The matrix elements of the \bt ~mode are given in
units of MeV$^{-1}$.}
\end{table}

For the case of the \bt ~decay mode, there are four active
configurations and each of them add coherently to the final matrix
element. They are partially suppressed by their amplitudes and the
final matrix element is of the order of 0.086 MeV $^{-1}$. This
value is about twice the one obtained by using the small
configuration space of subsection III.A. Consequently, the
calculated half-life, which is of the order of 1.68 $\times
{10}^{21}$ yrs, is shorter than the one obtained in the small
space. Concerning the \bz ~ decay mode, the results shown in Table
VI indicate that there is an interference between configurations
where both nucleons are in intruder orbits and those where the
proton occupies an intruder orbit. For this channel the resulting
matrix element is of the order of 0.293, a value which is about a
factor three smaller than the one obtained in the small
configuration space. The predicted half-life
\begin{equation}
\tau^{1/2}_{0\nu} ( \hbox{\gd} \rightarrow \hbox{\dy}) \cdot
\langle m_\nu \rangle^2
 = 2.05\times 10^{26} \hbox{yr},
\end{equation}
is an order of magnitude larger than the one obtained in the small
configuration space.

The above presented results can be summarized by noticing that the
effect of including occupations others than the most probable one
is less crucial for the two-neutrino mode than for the case of the
neutrinoless double beta decay. Nevertheless, the predicted
two-neutrino double beta decay of \gd ~ is still suppressed, as
compared to other double beta decay emitters. In this respect, the
results of the present calculations are an improvement of earlier
ones \cite{Cas94}, where claims about a suppression of the
two-neutrino mode have been made. Here, we have used a larger
configuration space, as explained before, instead of a single
configuration. The two neutrino double beta decay in \gd ~ is
hindered by nuclear structure effects, and the predicted half-life
is of the order of  $10^{21(22)}$ yr, depending upon the model
space. The zero neutrino double-beta-decay half-life is at least
three to four orders of magnitude larger. In view of these
predicted values, we are confident that the planned experiments
using GSO crystals \cite{Dan00} would definitely be able to detect
the \bt ~decay of \gd , and could establish competitive limits to
the \bz ~decay. The background suppression due to a large \bt
~half-life would be effective, although not as noticeably as was
optimistically envisioned in \cite{Dan00}.

Results about selection rules in other deformed double beta decay
emitters are reported in \cite{Hir02}.

\section{Conclusions}

In the present paper we have studied the \bt ~and \bz ~decay modes
of \gd ~to the ground states of \dy .
The transitions have been analyzed in the context of the pseudo SU(3) model.

The energy spectrum and electromagnetic transitions in \gd ~ and
\dy  ~ have been studied in detail, in previous works, using the
pseudo SU(3) model and a realistic Hamiltonian. Ground state wave
functions were built as linear combinations of the pseudo SU(3)
irreps associated with the larger quadrupole deformations, in a
model space with fixed occupation numbers in normal and unique
parity orbitals. Nucleons occupying intruder orbits were frozen.
The pseudo SU(3) leading irrep typically carries 60\% of the total
wave function.

In the present contribution the mixing of different occupation
numbers in the \dy ~ground state wave function was studied. Only
leading irreps, for each occupation, were considered in the
calculations. The mixing induced by the pairing interaction makes
possible the two neutrino double beta decay of \gd , which is
forbidden when only the most probable occupation numbers are used.

Explicit expressions are presented for the pairing mixing, and for
the \bt ~and \bz ~nuclear matrix elements in the present pseudo
SU(3) approach. The estimated \bb ~half-lives are larger than
those obtained using a spherical QRPA model, and the results
suggest that the planned experiments would succeed in detecting
the \bt ~decay in \gd , and in setting competitive limits for the
zero neutrino mode.

\section{Acknowledgment}

Work supported in part by CONACyT, and by a CONACyT-CONICET
agreement under the project {\em Algebraic methods in nuclear and
subnuclear physics.} O. Civitarese is a fellow of the CONICET,
Argentina.

\end{document}